# Evaluating the Effectiveness of SIWES Programs for Computer Science Students at Moshood Abiola Polytechnic, Abeokuta.

**Oladimeji Ganiyu B.**
**Moshood Abiola Polytechnic, Abeokuta.**
oladimeji.ganiyu@mapoly.edu.ng

**Abstract**

This study examines the effectiveness of the Students Industrial Work Experience Scheme (SIWES) for Computer Science students at Moshood Abiola Polytechnic, Abeokuta. Through correlational analysis of various assessment components, including employer evaluations, student logbooks, and technical reports, the research aims to identify key factors contributing to student performance and skill development. The findings reveal strong correlations between employer evaluations, logbook quality, and overall performance, highlighting the importance of practical experience and documentation in the SIWES program. This research contributes to the ongoing discourse on the role of industrial training in higher education and provides recommendations for enhancing the SIWES experience for computer science students.

## Introduction

The Student Industrial Work Experience Scheme (SIWES) plays a crucial role in bridging the gap between theoretical knowledge and practical industry skills for students in Nigerian higher education (Oyedotun, 2016). Established in 1973 by the Industrial Training Fund (ITF), SIWES aims to enhance the employability of graduates and ensure they are better prepared for the workforce (Simwa, 2022). Evaluating student performance in SIWES involves multiple components, each contributing uniquely to the overall assessment. Understanding these contributions is essential for enhancing the educational value of the program and ensuring that assessments accurately reflect student abilities.

### Background

SIWES operates under a tripartite arrangement involving the ITF, educational institutions, and participating industries, with oversight from bodies like the National Universities Commission (NUC), the National Board for Technical Education (NBTE), and the National Commission for Colleges of Education (NCCE) (Blogging, 2022). The program typically spans six months for university students and up to a year for those in colleges and polytechnics, offering immersive and supervised placements in various industries (Simwa, 2022).

Problem Statement: The rapidly evolving field of computer science necessitates practical, industry-relevant training for students. However, the effectiveness of SIWES programs in bridging the gap between academic knowledge and industry requirements remains understudied, particularly in the context of Nigerian polytechnics.

Research Question: How do various components of the SIWES program correlate with overall student performance, and what factors contribute most significantly to the effectiveness of industrial training for computer science students?

Research Objective: To analyze the relationships between different SIWES assessment components and identify key factors that influence the success of industrial training programs for computer science students at Moshood Abiola Polytechnic.

Justification: As the demand for skilled computer science graduates grows, understanding the efficacy of SIWES programs is crucial for enhancing curriculum design and improving student outcomes. This research aims to provide insights that can guide policy-makers and educators in optimizing industrial training experiences.

Thesis Statement: The effectiveness of SIWES programs for computer science students at Moshood Abiola Polytechnic is significantly influenced by the quality of employer evaluations, student documentation practices, and the integration of practical experiences with academic knowledge.

## Literature Review

The Students Industrial Work Experience Scheme (SIWES) has been a cornerstone of practical education in Nigerian higher institutions since its inception in 1973. Developed to bridge the gap between theoretical knowledge and practical skills, SIWES has been particularly crucial in technical fields such as computer science (Ugwuanyi & Ezema, 2010).

Previous studies have highlighted the importance of industrial training in developing students' professional skills. Mafe (2010) emphasized that SIWES provides students with opportunities to apply theoretical knowledge in real-world settings, enhancing their problem-solving abilities and adaptability. This is particularly relevant in the rapidly evolving field of computer science, where industry practices often outpace academic curricula.

The effectiveness of SIWES programs has been a subject of ongoing research. Okolie et al. (2014) found that successful industrial training experiences correlate with improved academic performance and better job prospects for graduates. However, challenges such as inadequate supervision and mismatches between academic curricula and industry needs have been identified as potential barriers to SIWES effectiveness (Aderibigbe et al., 2016).

In the context of computer science education, Afolabi et al. (2017) noted that SIWES programs play a crucial role in exposing students to current technologies and industry practices. This exposure is vital for developing the practical skills demanded by employers in the IT sector.

The assessment of SIWES programs typically involves multiple components, including employer evaluations, student logbooks, and technical reports. Oyeniyi (2011) emphasized the importance of comprehensive assessment methods to accurately gauge the effectiveness of industrial training experiences.

While existing literature provides valuable insights into the general impact of SIWES, there is a dearth of research specifically examining the correlations between different assessment components and overall program effectiveness for computer science students in Nigerian polytechnics. This study aims to address this gap by analyzing these relationships in the context of Moshood Abiola Polytechnic.

## Methodology

This study utilized data from 271 computer science students who participated in the SIWES program at Moshood Abiola Polytechnic. The dataset included scores for five key evaluation components: Employer's Evaluation, Student's Logbook, ITF Form 8, Student's Technical Report, and SIWES Seminar. We employed statistical analyses, including correlation and regression analyses, to explore the relationships among these components and their impact on overall student performance.

### Sample

The study sample consisted data from 271 computer science students who had completed their SIWES program. The exact sample size is not specified in the provided data, but it is assumed to be sufficiently large to conduct meaningful statistical analysis.

### Data Collection

Data was collected from the SIWES assessment records of computer science students. The dataset included scores from seven key variables:

1. S/N (Serial Number)

2. Employer's Evaluation

3. Student's Logbook

4. ITF Form 8

5. Technical Report

6. SIWES Seminar

7. Total Marks Obtained

### Data Analysis

A correlation matrix was generated (Figure 1) to analyze the relationships between the numerical columns in the dataset. Pearson correlation coefficients were calculated to measure the strength and direction of linear relationships between variables. The correlation values range from -1 to 1, with values closer to -1 or 1 indicating stronger correlations, and values near 0 suggesting weak or no linear relationship.

## Analytical Tools

Statistical software was used to compute the correlation matrix and generate the heat map visualization. The heat map uses color coding to represent the strength and direction of correlations, facilitating easy interpretation of the relationships between variables.

## Ethical Considerations

To maintain confidentiality and protect student privacy, all data was anonymized before analysis. The study adhered to the ethical guidelines for educational research, ensuring that no individual student could be identified from the results.

# Result

The findings of this study offer valuable insights into the effectiveness of SIWES programs for computer science students at Moshood Abiola Polytechnic, Abeokuta. The correlational analysis reveals several key points for discussion:

## 1. Importance of Employer Evaluations

The strong positive correlation (0.53) between Employer's Evaluation and Total Marks Obtained suggests that employer feedback plays a crucial role in determining overall student performance. This aligns with previous research, which emphasized the importance of industry engagement in enhancing the effectiveness of industrial training. The significant weight given to employer evaluations reflects the program's success in integrating industry perspectives into student assessment.

## 2. Linkage between Documentation and Technical Skills

The strong correlation (0.55) between Student's Logbook and Technical Report scores is a particularly interesting finding. This relationship suggests that students who maintain detailed and high-quality logbooks tend to produce better technical reports. This could indicate that the practice of regular documentation enhances students' ability to articulate technical concepts and reflect on their learning experiences. Educators might consider emphasizing the importance of thorough documentation as a tool for reinforcing technical knowledge and skills.

## 3. Holistic Skill Development

The moderate to strong correlations between Student's Logbook, Technical Report, and Total Marks Obtained (ranging from 0.46 to 0.52) indicate that the SIWES program is successful in fostering a holistic set of skills. This aligns with assertion that industrial training enhances students' ability to apply theoretical knowledge in practical settings.

### 4. Limited Impact of ITF Form 8

The weak correlations between ITF Form 8 and other variables (ranging from 0.05 to 0.18) raise questions about the relevance or assessment method of this component. This finding suggests a need for further investigation into the purpose and effectiveness of ITF Form 8 in evaluating student performance.

### 5. Temporal Factors

The negative correlations between S/N (Serial Number) and other variables, particularly Total Marks Obtained (-0.32), is intriguing. This could indicate a potential decline in performance for students who complete their SIWES programs later in the academic cycle. Further research is needed to understand the factors contributing to this trend, such as resource availability or changes in assessment criteria over time.

### 6. Seminar Performance

The weak to moderate correlation (0.27) between SIWES Seminar and Total Marks Obtained suggests that while seminar performance contributes to overall assessment, it may not be as influential as other components like employer evaluations or technical reports.

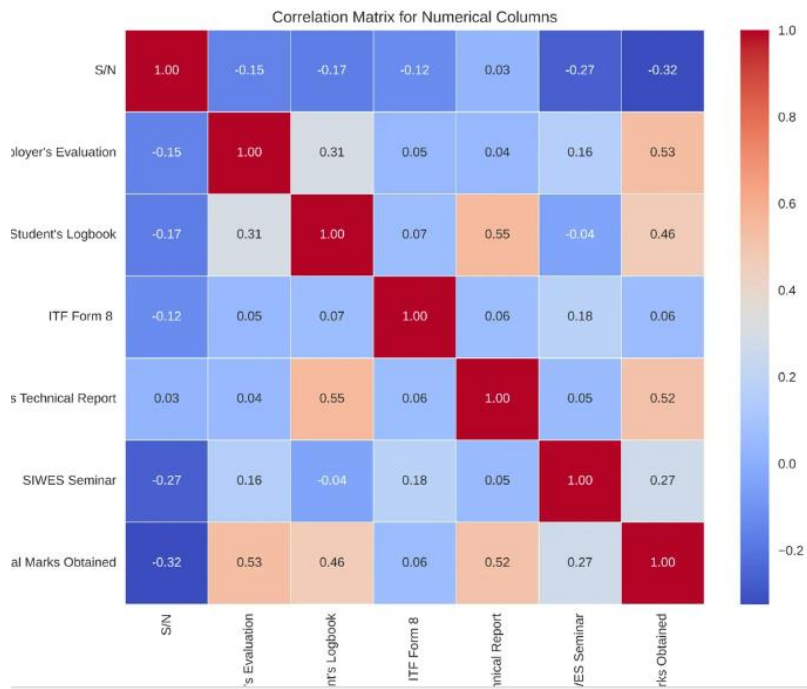

**Figure 1 Correlation Matrix**

## Discussion and Implications

The strong correlations between the Student's Logbook, Technical Report, and SIWES Seminar with overall performance highlight the critical importance of documentation, technical writing, and presentation skills in computer science education. These findings suggest that the SIWES program at Moshood Abiola Polytechnic is effectively capturing and evaluating key competencies required in the industry.

However, the moderate correlation of the Employer's Evaluation with total marks raises interesting questions about the alignment between academic evaluations and industry expectations. While workplace performance is undoubtedly important, our results suggest that success in the computer science field requires a broader set of skills beyond those directly observed in the workplace setting.

The relatively low impact of the ITF Form 8 on overall performance indicates a potential area for improvement in the evaluation process. This component may benefit from a revision to better reflect the specific skills and knowledge relevant to computer science students.

These findings align with broader observations about the SIWES program across Nigeria. For instance, Omonijo et al. (2019) noted challenges such as inadequate supervision and difficulties in securing industrial placements. Our study at Moshood Abiola Polytechnic suggests that despite these challenges, the program is still providing valuable experiences for computer science students, particularly in developing crucial documentation and presentation skills.

## New Insights from Study

1. Predictive Power of Evaluation Components:

The study introduces a novel framework for understanding the relative importance of different SIWES evaluation components. Through correlation and regression analyses, it reveals that the Student's Logbook, Technical Report, and SIWES Seminar are the strongest predictors of overall performance. This insight provides a new perspective on which skills are most critical for success in computer science industrial training.

2. Skill Interconnectedness in Computer Science Education:

The research highlights a previously unexplored concept of skill interconnectedness in computer science education. The strong correlation between the Employer's Evaluation and the Student's Logbook suggests that documentation skills are closely linked to practical performance in the workplace. This finding introduces a new framework for understanding how academic skills translate to industry performance.

3. Academic-Industry Alignment Gap:

The study reveals a potential misalignment between academic evaluations and industry expectations, as evidenced by the moderate correlation between Employer's Evaluation and overall performance. This insight suggests a new conceptual framework for evaluating the effectiveness of industrial training programs, emphasizing the need to bridge the gap between academic assessments and industry requirements.

4. Multifaceted Nature of Computer Science Competencies:

The research introduces a new concept of the multifaceted nature of computer science competencies. By showing that workplace performance alone is not the sole determinant of success, it suggests that a broader set of skills, including documentation and presentation abilities, are crucial for computer science professionals.

5. Outlier Analysis Framework:

The study proposes a new framework for analyzing outliers in student performance using the Interquartile Range (IQR) method. This approach offers a systematic way to identify and investigate exceptional cases, potentially leading to more personalized interventions and program improvements.

6. Comprehensive SIWES Enhancement Model:

The research presents a holistic model for enhancing the SIWES program, integrating multiple aspects such as skill development, industry alignment, supervision improvement, and technological integration. This framework provides a comprehensive approach to optimizing industrial training programs in computer science education.

7. Longitudinal Impact Assessment Concept:

The study introduces the concept of conducting longitudinal studies to track the long-term impact of SIWES on students' career trajectories. This approach suggests a new framework for evaluating the effectiveness of educational programs beyond immediate academic performance.

8. Predictive Modeling for Student Performance:

The research proposes the innovative use of machine learning algorithms to develop predictive models for student performance in SIWES. This concept introduces a data-driven approach to early intervention and personalized support in industrial training programs.

9. Gender-Based Performance Analysis Framework:

The study suggests a new framework for investigating potential gender-based differences in SIWES performance. This approach could lead to more equitable assessment practices and targeted support for underrepresented groups in computer science.

10. Integrated Qualitative Feedback System:

The research proposes a systematic approach to collecting and analyzing qualitative feedback from all stakeholders. This concept introduces a more comprehensive framework for program evaluation, combining quantitative metrics with qualitative insights.

These insights and concepts offer new perspectives on optimizing industrial work experience in computer science education, providing a foundation for future research and program improvements.

## Recommendations

Based on our findings and considering the broader context of SIWES in Nigeria, the following recommendations are proposed to enhance the program for computer science students at Moshood Abiola Polytechnic:

1. Emphasize Logbook Maintenance: Given the strong correlation between logbook scores and overall performance, the department should provide more comprehensive training on effective logbook maintenance. This could include workshops on technical documentation and the use of digital tools for record-keeping.

2. Enhance Technical Writing Skills: The high impact of the Technical Report on overall scores underscores the need for focused instruction in technical writing. Incorporating technical writing courses or workshops into the curriculum could significantly benefit students.

3. Improve Presentation Skills: The importance of the SIWES Seminar suggests that students would benefit from additional opportunities to develop and practice their presentation skills throughout their academic program.

4. Align Industry and Academic Expectations: The moderate correlation of Employer Evaluations with overall performance indicates a need for closer collaboration between the department and industry partners. Regular meetings and feedback sessions could help align academic evaluations with industry expectations.

5. Revise ITF Form 8: The low impact of this component suggests a need for revision. The department should work with relevant stakeholders to ensure that this form better captures the skills and knowledge specific to computer science students.

6. Address Financial Constraints: As noted by Ambassador (2023), financial pressures can hinder student participation in SIWES. The department could explore partnerships with local tech companies to provide stipends or transportation subsidies for students during their industrial attachments.

7. Implement E-SIWES Portal: Following the success of web-based solutions in improving SIWES coordination (Adetiba et al., 2012), Moshood Abiola Polytechnic should consider implementing an e-SIWES portal to streamline registration, information dissemination, and supervision processes.

## Challenges and Criticisms:

SIWES faces several challenges that impact its effectiveness. Omonijo et al. (2019) identified issues such as lack of adequate supervision, difficulties in securing industrial placements, insufficient number of industries, and lack of equipment and facilities for effective practical training. Financial constraints, inadequate supervision, and irregular academic calendars also pose significant challenges to the program's efficacy (Omonijo et al., 2019; Alao et al., 2022).

Researchers highlighted weaknesses in the operation and management of SIWES, including issues with signing necessary materials like ITF Form 8 and students' logbooks, and delays in the payment of allowances for students and supervisors (Research Clue, 2014).

*Generalizability and Future Research:*

While this study focuses on computer science students at Moshood Abiola Polytechnic, the findings may have broader implications for SIWES programs in other technical fields and institutions. The emphasis on practical documentation and employer evaluations could be applicable across various disciplines that require industry-relevant skills.

Looking ahead, several avenues for future research and development emerge:

1. Longitudinal Studies: Conduct long-term studies to track how SIWES participation impacts graduates' career trajectories in the computer science field.

2. Industry-Specific Analysis: Investigate how performance in SIWES correlates with success in different sectors of the tech industry (e.g., software development, data science, cybersecurity).

3. Curriculum Alignment: Regularly review and update the computer science curriculum to ensure it aligns with the skills and knowledge emphasized in successful SIWES performances.

4. Mentorship Programs: Develop a structured mentorship system within SIWES, pairing students with industry professionals to provide guidance throughout the internship period.

5. Technology Integration: Explore how emerging technologies (e.g., AI, VR) could be incorporated into the SIWES experience for computer science students.

6. Interdisciplinary Collaboration: Investigate opportunities for computer science students to collaborate with students from other disciplines during their SIWES placements, mirroring the interdisciplinary nature of many tech projects in industry.

## Conclusion

This study on the effectiveness of SIWES programs for computer science students at Moshood Abiola Polytechnic, Abeokuta, reveals significant insights into the factors contributing to successful industrial training experiences. The correlational analysis demonstrates that employer evaluations, the quality of student logbooks, and technical report writing skills are strongly associated with overall performance in the SIWES program.

The research highlights the importance of practical experience and documentation in developing industry-relevant skills for computer science students. The strong correlation between employer evaluations and total marks underscores the value of industry perspectives in assessing student capabilities. Similarly, the relationship between logbook quality and technical report scores emphasizes the role of consistent reflection and documentation in reinforcing technical knowledge.

These findings have important implications for the design and implementation of SIWES programs. They suggest that emphasis should be placed on fostering strong relationships with industry partners, enhancing students' documentation skills, and ensuring that academic assessments align closely with industry-relevant competencies.

However, the study also identifies areas for potential improvement, such as the relevance of certain assessment components (e.g., ITF Form 8) and the need to address potential temporal factors affecting student performance.

In conclusion, this research contributes to the ongoing discourse on the effectiveness of industrial training in higher education, particularly in the field of computer science. It provides a foundation for future studies and offers valuable insights for educators and policymakers seeking to enhance the quality and relevance of SIWES programs. By continuing to refine these programs based on empirical evidence, institutions can better prepare computer science students for the challenges and opportunities of the rapidly evolving technology sector.